\documentstyle{article}
\date{\today}

\begin{document}
\title{$D^{\ast}\to D\gamma$ and $B^{\ast}\to B\gamma$ as derived from 
QCD Sum Rules}

\author{{Shi-lin Zhu,$^1$ W-Y. P. Hwang,$^{2,3}$ and Ze-sen Yang$^1$}\\
{$^1$Department of Physics, Peking University, Beijing, 100871, China}\\
{$^2$Department of Physics, National Taiwan University, Taipei, 
Taiwan 10764}\\
{$^3$Center for Theoretical Physics, Laboratory for Nuclear Science 
       and}\\
{Department of Physics, Massachusetts Institute of Technology,}\\ 
{Cambridge, Massachusetts 02139}
}
\maketitle

\begin{center}
\begin{minipage}{140mm}
\vskip 0.5 cm
\begin{center}{\bf Abstract}\end{center}
{\large
The method of QCD sum rules in the presence of the external electromagnetic 
$F_{\mu\nu}$ field is used to analyze radiative decays of charmed or 
bottomed mesons such as $D^{\ast}\to D\gamma$ and $B^{\ast}\to B\gamma$, 
with the susceptibilities obtained previously from the study of baryon  
magnetic moments. Our predictions on $D^{\ast}$ decays agree very well with 
the experimental data. There are differences among the various theoretical 
predictions on $B^{\ast}$ decays but the data are not yet available.
\vskip 0.5cm
PACS Indices: 14.40.Jz; 13.20.Fc; 13.20.Jf
}
\end{minipage}
\end{center}

\newpage
\large

\par
Heavy flavor physics provides a natural and convenient laboratory for
testing basic ideas such as the CKM matrix, heavy quark symmetry, and 
aspects
related to quantum chromodynamics (QCD). Indeed, it is often believed that
the nonperturbative aspects of QCD may play a less significant role 
in the heavy quark sector as compared to that in the light quark systems, 
thereby offering simplifications in testing fundamental ideas.

The situation regarding radiative decays of charmed and bottomed mesons 
such as $D^{\ast}\to D\gamma$ and $B^{\ast}\to B\gamma$ is a little 
confusing at the moment. In the charmed sector, the predictions [1,2] based 
on heavy quark symmetry differ from those [3] obtained by implementing 
simultaneously both heavy quark symmetry and chiral symmetry. And, yet, 
these predictions also differ significantly from those \cite{JAUS} 
obtained in a relativistic constituent quark model. The disparity 
among these predictions
makes it very difficult, if not impossible, to assess the significance of
the underlying idea(s) under investigation.
 
As a potentially useful benchmark, we wish to present and discuss 
the predictions which may be obtained by employing the method of QCD sum
rules \cite{SVZ}, especially using such method in the presence of an   
external electromagnetic field \cite{IOFFE, Balit}. The important aspect 
here is that the various susceptibilities involved in such method have 
already been determined from previous studies of baryon magnetic moments 
and as a result our predictions are parameter-free on top of the 
straightforward nature of applying the method to the heavy quark sector.

\par
We recall \cite{IOFFE,Balit} that the 
two-point correlation function $\Pi(p)$ in the presence of the constant 
electromagnetic field $F_{\mu\nu}$ may be used as the basis for the 
derivation of QCD sum rules. Using the radiative decay ${B^{\ast}}^0 \to 
B^0 \gamma $ as a specific example, we have 
\begin{equation}
\label{33}
\Pi_\mu (p)=i\int d^4x e^{ipx}\langle 0|T[j^{{B^{\ast}}^0}_\mu(x),
j^{B^0}(0)] |0\rangle_{F_{\alpha\beta}} \, ,
\end{equation}
where the interpolating fields are
\begin{equation}
\begin{array}{lll}
j_\mu^{{B^{\ast}}^0}(x)={\overline b}^a(x)\gamma_{\mu} d^a(x) \, , &   &
j^{B^0}(x)={\overline b}^a(x)i\gamma_5 d^a(x) \, ,
\end{array}
\end{equation}
with $a$ the color index. The decay constants $f_{B^{\ast}}$ and $f_B$ 
are defined as 
\begin{equation}
\begin{array}{c}
\langle 0|j_\mu^{B^{\ast}}(0)|B^{\ast}(p) \rangle =f_{B^{\ast}}\,
m_{B^{\ast}}\, \epsilon_{\mu} \, , \\
\langle 0|j^B(0)|B (p) \rangle =f_B {{m^2_B} \over {m_b +m_d}} \, ,
\end{array}
\end{equation}
where $\epsilon_{\mu}$ is the polarization 4-vector of the $B^\ast$ meson.

\par
The decay process $B^\ast(1^-) \to B (0^-)+ \gamma$ may be characterized 
uniquely as follows:
\begin{equation}
{\cal L}_{B^{\ast}B\gamma}=e {{g_{B^{\ast}B\gamma}}\over {m_{B^{\ast}}} }
\epsilon_{\mu\nu\lambda\delta}
\partial^{\nu}A^{\mu} {B^{\ast}}^{\lambda} \partial^{\delta} B \, ,
\end{equation}
where $A_{\mu}$ is the electromagnetic field. The corresponding partial width 
is
\begin{equation}
\Gamma(B^{\ast}\to B\gamma)=\alpha {{g^2_{B^{\ast}\to B\gamma}}\over {3} }
{{P_{\mbox{cm}}^3}\over {m_{B^{\ast}}^2}}  \, ,
\end{equation}
with $P_{\mbox{cm}}$ the $B$ meson momentum in the CM frame (i.e. in the 
rest frame of the initial vector meson $B^\ast$).

\par
The external field $F_{\mu\nu}$ may induce changes in the physical vacuum and 
modify the propagation of quarks. Up to dimension six ($d\leq 6$), we may
introduce three induced condensates as follows.
\begin{equation} 
\begin{array}{c}
\langle 0 | {\overline  q} \sigma_{\mu\nu} q |0 \rangle_{F_{\mu\nu}} = e_q e \chi
F_{\mu\nu} \langle 0 | {\overline  q}  q |0 \rangle \, , \\ 
g_c \langle 0 | {\overline  q} {{\lambda^n}\over {2}}G^n_{\mu\nu} q |0 \rangle_{F_{\mu\nu}}
= e_q e \kappa F_{\mu\nu} 
\langle 0 | {\overline  q}  q |0 \rangle \, , \\
g_c \epsilon^{\mu\nu\lambda\sigma} 
\langle 0 | {\overline  q} \gamma_5 {{\lambda^n}\over {2}} G^n_{\lambda\sigma} q |0 \rangle_{F_{\mu\nu}}
= i e_q e \xi F^{\mu\nu} 
\langle 0 | {\overline  q}  q |0 \rangle \, ,
\end{array}
\end{equation}
where $q$ refers mainly to $u$ or $d$ with suitable modifications for the $s$ 
quark (while the condensates involving the charm or bottom quark are assumed 
to be negligible), $e$ is the unit charge, $e_u= {2 \over 3}$, and 
$e_d=-{1\over 3}$. In
studying baryon magnetic moments, Ioffe and Smilga \cite{IOFFE} obtained the 
nucleon anomalous magnetic moments $\mu_p =3.0$ and  
$\mu_n=-2.0$ $(\pm 10\%)$ with the quark condensate susceptibilities 
$\chi \approx -8\,\mbox{GeV}^2$ and $\kappa$, $\xi$ quite small. 
Balitsky and Yung \cite{Balit} adopted the one-pole approximation to estimate
the susceptibilities and obtained,
\begin{equation}
\begin{array}{lll}
\chi=-3.3\,\mbox{GeV}^{-2},& \kappa=0.22,& \xi=-0.44 \, .
\end{array}
\end{equation}
Subsequently Belyaev and Kogan \cite{Kogan} extended the calculation and 
obtained an improved estimate $\chi =-5.7\,\mbox{GeV}^{-2}$ using the two-pole 
approximation. Chiu et al \cite{Chiu} also estimated the susceptibilities 
with two-pole model and obtained
\begin{equation}
\begin{array}{lll}
\chi=-4.4\,\mbox{GeV}^{-2},& \kappa=0.4,& \xi=-0.8 \, .
\end{array}
\end{equation}
Furthermore, Chiu and co-workers \cite{Chiu} re-analysed the various sum 
rules by treating $\chi$, $\kappa$, and $\xi$ as free parameters to provide
an overall fit to the observed baryon magnetic moments.
The opitimal values which they obtained are given by 
\begin{equation}
\begin{array}{lll}
\chi=-3\,\mbox{GeV}^{-2},& \kappa=0.75,& \xi=-1.5 \, .
\end{array}
\end{equation}
We note that in these analyses the susceptibility values are consistent 
with one another except for the earliest result $\chi=-8\,\mbox{GeV}^{-2}$ 
in \cite{IOFFE}, which is considerably larger (in magnitude). In what 
follows, we shall adopt the condensate parameters $\chi=-3.5\,
\mbox{GeV}^{-2}$, 
$2\kappa +\xi \approx 0$ with $\kappa$, $\xi$ quite small, which represent the 
average of the latter values discussed above. The same set of susceptibilities 
yields a very good overall agreement \cite{Hwang} with the experiment for the 
vector meson decays in $SU_f(3)$ sector $(u,\, d,\, s)$.
\par
At the quark level, the correlation function $\Pi_{\mu} (p)$ may be evaluated  
with the aid of the formula:
\begin{equation}
\Pi_{\mu} (p)= -i\int d^4x e^{ipx} Tr [ iS^{ef}_b(-x)\gamma_{\mu} 
iS_d^{fe}(x) i\gamma_5]
\end{equation}
where $iS^{ab}(x)$ is the quark propagator in the presence of the external 
eletromagnetic field $F_{\mu\nu}$ \cite{IOFFE}. To incorporate the (heavy) 
quark mass, we choose to work in momentum space. For example, we have, for  
the down quark propagator,
\begin{equation}
\begin{array}{ll}
iS^{ab}_d(p)=
& \delta^{ab}  {{i} \over { {\hat p} -m_d}} \\
&+ {i\over 4} {\lambda_{ab}^n \over 2} g_c G_{\mu\nu}^n {1\over (p^2-m_d^2)^2}
       [\sigma^{\mu\nu} ({\hat p}+m_d) +({\hat p}+m_d) \sigma^{\mu\nu}) ] \\
&+ {i\over 4} e_d \delta^{ab} F_{\mu\nu}  {1\over (p^2-m_d^2)^2}
       [\sigma^{\mu\nu} ({\hat p}+m_d) +({\hat p}+m_d) \sigma^{\mu\nu})  ]  \\
&-\delta^{ab}  { \langle {\overline d}d \rangle \over 12} (2\pi )^4\delta^4(p)\\
&-\delta^{ab}  { \langle g_c {\overline d} \sigma \cdot G d \rangle \over 192} 
  (2\pi )^4 g^{\mu\nu} \partial_\mu \partial_\nu \delta^4(p) \\
&-\delta^{ab}e_d  { \langle {\overline d}d \rangle \over 192} 
  [ \sigma\cdot F g^{\mu\nu} - {1\over 3} \gamma^{\mu} \sigma\cdot F 
  \gamma^{\nu} ] (2\pi )^4 \partial_\mu \partial_\nu \delta^4(p)  \\
&-\delta^{ab} e_d \chi  { \langle {\overline d} d \rangle \over 24} 
  \sigma\cdot F(2\pi )^4\delta^4(p)                               \\
&-\delta^{ab}e_d \kappa  { \langle  {\overline d} d \rangle \over 192} 
  [ \sigma \cdot F g^{\mu\nu}- {1\over 3}\gamma^{\mu}
  \sigma \cdot F \gamma^{\nu} ]
  (2\pi )^4 \partial_\mu \partial_\nu \delta^4(p)                   \\
&+i \delta^{ab}e_d \xi  { \langle  {\overline d} d \rangle \over {2^8 \times 3}} 
  [ \sigma^{\delta\rho}  g^{\mu\nu}- {1\over 3}\gamma^{\mu} 
  \sigma^{\delta\rho}  \gamma^{\nu} ]\gamma_5 \epsilon_{\alpha\beta\delta\rho} F^{\alpha\beta}
  (2\pi )^4 \partial_\mu \partial_\nu \delta^4(p)                     \\
&+ \cdots,
\end{array}
\end{equation}
with ${\hat p}\equiv p_\mu \gamma^\mu$. Here we follow \cite{IOFFE,Balit} 
and do not introduce induced condensates of higher dimensions. 
\par
Substituting the quark propagator back into $\Pi_\mu(p)$ (and neglecting  
condensates related to the heavy quark), we obtain
\begin{equation}
\begin{array}{ll}
\label{lhs}
\Pi_{\mu}^{l.h.s.}(p)=&\{  {3 \over {8\pi^2}}m_b \int_0^1 dx  {{e_b (1-x)+e_d x}
 \over {m_b^2-p^2 x}}
 - {e_d \over 2}\chi \langle 0 | {\overline  d}  d |0 \rangle  {{1}\over {p^2 -m^2_b}} \\
& + {{e_d -e_b-\xi }\over{2}} \langle 0 | {\overline  d}  d |0 \rangle  {{1} \over {(p^2 -m^2_b)^2}} 
+ {e_d \over 3} (\xi -\kappa-1) \langle 0 | {\overline  d}  d |0 \rangle  {{p^2}\over {(p^2 -m^2_b)^3}} \\
&+ {e_b \over 24}\langle 0 |g_c {\overline  d} \sigma\cdot G d |0 \rangle
 {{1}\over {(p^2 -m^2_b)^3}  }
-( {3\over 8}e_b + {e_d \over 4})\langle 0 |g_c {\overline  d} \sigma\cdot G d |0 \rangle
 {{m^2}\over {(p^2 -m^2_b)^4}}  \\
&+\cdots \}\epsilon_{\alpha\beta\mu\sigma} F^{\alpha\beta} p^{\sigma} 
\end{array}
\end{equation}
Note that the operators with lowest dimension ($d=1$) arise from the heavy 
quark mass $m_h$ and the induced light quark condensate 
$\chi \langle 0 | {\overline  q}  q |0 \rangle $, which are respectively
the first two terms in Eq. (\ref{lhs}). The next two terms represent 
dimension-three contributions arising from the three condensates 
$\langle 0 | {\overline  q}  q |0 \rangle$, 
$\kappa\langle 0 | {\overline  q}  q |0 \rangle$, and 
$\xi\langle 0 | {\overline  q}  q |0 \rangle$. 
At dimension five, we have kept only the quark-gluon mixed condensate 
$[$the remaining two terms in (\ref{lhs})$]$ as the contributions from gluon
condensates are found to be negligible.

\par 
The short-distance operator-product expansion (\ref{lhs}) converges rapidly. 
For bottomed mesons, the perturbative term contributes up to as much as 
$90\%$ of the total sum rules (at $p^2=0$), while the induced quark 
condensate dominates the power corrections and its contribution is about 
$9\%$. The contributions from the condensates of higher dimensions 
cancel among themselves and the net results are very small. For charmed 
mesons, the perturbative contribution is about $70\%$ as due to the 
comparatively smaller charm quark mass. Again the contribution from 
the induced quark condensate is dominant among the remaning power corrections. 
We wish to point out that, unlike in the QCD sum rules analyses of charmonium 
where the perturbative contribution and the gluon condensate play a very 
important role, in the present analysis the contributions from the light 
quark condensates $\chi \langle 0 | {\overline  q}  q |0 \rangle $ and 
$\langle 0 | {\overline  q}  q |0 \rangle $ are in fact much more 
important than that from the gluon condensate. 
 
\par
At the hadronic level, we express $\Pi(p)$ as follows:
\begin{equation}
\label{rhs}
\label{7}
\Pi_{\mu}^{r.h.s.}(p)=f_{B^{\ast}}f_B  {m_B^2  \over  m_b+m_d} { g_{B^{\ast}B\gamma}   \over  2}
 {1  \over  p^2 -m_{B^{\ast}}^2}
 {1  \over  p^2 -m_B^2} \epsilon_{\alpha\beta\mu\sigma} F^{\alpha\beta} p^{\sigma}
+\cdots \, ,
\end{equation}
where we write down explicitly the leading $B^{\ast}\to B\gamma$ 
contribution and denote the contribution from the excited states and 
continuum simply by the ellipsis.

\par
Unlike for light quark systems, Borel transform is no longer the most 
efficient way to accentuate the contribution from the ground state. The 
reason is twofold:  First of all, we note that in performing Borel 
transform the single-pole contributions from excited states give rise to 
an additional term $AM_B^2$ on the phenomenological side of the sum 
rule \cite{IOFFE,Be95}. Due to the small energy gap between the ground 
state and excited states in heavy meson systems, this term could be fairly 
large. On the other hand, it was suggested \cite{SVZ,REINDERS} that a method 
of taking suitable moments (as used below) could be more efficient to 
suppress the contributions from the excited states, as the heavy quark mass 
provides a natural scale ($m_h \gg \Lambda_{QCD}$).

\par
To simplify the numerical manipulation we adopt the following approximation:
\begin{equation}
 {1  \over  p^2 -m_{B^{\ast}}^2} {1  \over  p^2 -m_B^2} \approx 
 {1  \over  (p^2- {\overline m}_B^2)^2},
\end{equation}
where ${\overline m}_B^2= {m_{B^{\ast}}^2+m_B^2  \over  2}$. The approximation
is justified to the level of one part in $10^4$ (since $m_{B^{\ast}} -m_B 
=46\,\mbox{MeV}$ and we are very far away from either of the two poles). 

\par
In order to pick out the ground state contribution we define moments by taking
derivatives of the correlator $\Pi(p)$ (the coefficient of $\epsilon_{\alpha
\beta\mu\sigma}F^{\alpha\beta}p^\sigma$ in $\Pi_\mu(p)$) 
\cite{SVZ,REINDERS},
\begin{equation}
\begin{array}{ll}
M_n\equiv  {1  \over  (n+1)!}( {d  \over  dp^2})^n \Pi(p)|_{p^2=-Q^2_0},\qquad
& (Q^2_0 \geq 0).
\end{array}
\end{equation}
At the phenomenological side (r.h.s.) we have
\begin{equation}
M_n^{r.h.s.}(Q^2_0)=f_{B^{\ast}}f_B  {m_B^2  \over  m_b+m_d} {g_{B^{\ast}B\gamma}  \over  2}
 {1  \over  ({\overline m}_B^2+Q^2_0)^{n+2}} [1+\delta_n(Q^2_0)]
\end{equation}
where $\delta_n(Q^2_0)$ contains the contributions from the higher resonances 
and the continuum. For high n, $\delta_n(Q^2_0)$ goes to zero rapidly because
of the factors $( {{\overline m}_B^2+Q_0^2  \over  
{\overline m}_{B^{\prime}}^2+Q_0^2})^{n+2}$ 
it contains and because ${\overline m}_{B^{\prime}} > {\overline m}_B$. So 
from a certain $n$ onwards $M_n^{r.h.s.}(Q^2_0)$ will be practically equal to 
the contribution of the ground state. We also introduce the ratios of the 
moments:
\begin{equation}
r_n(Q_0^2)\equiv  {M_{n-1}(Q_0^2)  \over  M_n(Q_0^2)}=({\overline m}_B^2+Q_0^2)
 {1+\delta_{n-1}(Q_0^2)  \over  1+\delta_n(Q_0^2)}\, ,
\end{equation}
which immediately yields the mass ${\overline m}_B^2$ if we are at 
sufficiently 
high $n$ where $\delta_{n}(Q_0^2)\approx \delta_{n-1}(Q_0^2)$ \cite{REINDERS}. 
We use the ratios to check the convergency and consistency of our sum rule. As
observed in \cite{SVZ,REINDERS} the stability region in $n$ will change with 
the parameter $y= {Q_0^2  \over  m_b^2}$ and one should study the stability of the 
moments for growing $y$. A plateau develops for certain $n$ and $y$ if the sum 
rules work well, which arises from the balance of the pertuabative 
contributions 
and power corrections. We have checked that, in the region $y=0.3\sim 0.7$ 
and $n=11\sim 16$, the mass ${\overline m}_B =
\sqrt{ {m_{B^{\prime}}^2 +m_B^2  \over  2}}$ extracted in this way agrees very 
well with the experimental value. The heavy quark mass is scale dependent. 
We take $m_b =4.7\,\mbox{GeV}$ and $m_c =1.3\,\mbox{GeV}$ as commonly adopted 
in the QCD sum rule approach \cite{SVZ,Be95}. We take the values from the 
standard two-point sum rules for the decay constants $f_B$ and 
$f_{B^{\ast}}$ \cite{Be95}
\begin{equation}
\begin{array}{ll}
f_B=140\,\mbox{MeV}\, , &f_{B^{\ast}}=160\,\mbox{MeV}\, .
\end{array}
\end{equation}
Using the experimental heavy meson masses $m_B=5.28\,\mbox{GeV}$ and 
$m_{B^{\ast}}= 5.33\,\mbox{GeV}$ \cite{DATA} and equating 
\begin{equation}
M_n^{l.h.s.}(Q^2_0)=M_n^{r.h.s.}(Q^2_0)
\end{equation}
with $y$, $n$ in 
the region $y=0.3\sim 0.7$, $n=11\sim 16$ 
where the plateau develops, we finally obtain 
\begin{equation}
\begin{array}{c}
g_{{B^{\ast}}^0 B^0 \gamma} =-4.0 \pm 0.4 \, , \\
g_{{B^{\ast}}^+ B^+ \gamma} =6.8 \pm 0.6 \, ,    \\
g_{B_s^{\ast} B_s \gamma} =-5.0 \pm 0.5 \, .
\end{array}
\end{equation}
We have included the correction due to the strange quark mass $m_s=150\,
\mbox{MeV}$ which reproduces roughly the $B_s^{\ast}-B^{\ast}$ mass 
splitting. Note that the indicated errors are due to the method of QCD 
sum rules 
itself. Typically in the region $y=0.3\sim 0.7$ and $n=11\sim 16$ the 
perturbative term, the induced quark condensate, the dimension-three 
correction, and the dimension-five correction contribute $56\%$, $57\%$, 
$-18\%$, and $5\%$ of the sum rule, respectively. Using the above couplings, 
we obtain the partial width as follows:
\begin{equation}
\begin{array}{c}
\Gamma_{{B^{\ast}}^0\to B^0 \gamma} =0.13\pm 0.03 \mbox{keV}\, , \\
\Gamma_{{B^{\ast}}^+\to B^+ \gamma} =0.38\pm 0.06 \mbox{keV}\, , \\
\Gamma_{B_s^{\ast}\to B_s \gamma} =0.22\pm 0.04 \mbox{keV}\, ,
\end{array}
\end{equation}
which are also listed in Table I. The experiments to measure these partial 
widths are yet to be carried out.

\par
The above derivation holds for charmed mesons, except that 
the charm quark is not so heavy as the bottom quark so that 
the plateau develops at larger $y$ ($y = 1.5 \sim 2.0$, $n = 5 \sim 8$).
Using the sum rule values $f_D=170\pm 10\,\mbox{MeV}$ and 
$f_{D^{\ast}}=240\pm 20\, \mbox{MeV}$ and the observed masses 
$m_D=1.87\, \mbox{GeV}$ and $m_{D^{\ast}}=2.01\mbox{GeV}$, we obtain the 
couplings: 
\begin{equation}
\begin{array}{c}
g_{{D^{\ast}}^0 D^0 \gamma} =2.88 \pm 0.3 \, , \\
g_{{D^{\ast}}^+ D^+ \gamma} =-0.38 \pm 0.1 \, ,  \\
g_{D_s^{\ast} D_s \gamma} =-0.3 \pm 0.1 \, .
\end{array}
\end{equation}
The sum rule for $g_{{D^{\ast}}^0 D^0 \gamma}$ is stable. However, the 
sum rule for $g_{{D^{\ast}}^+ D^+ \gamma}$ suffers from the fact that  
the perturbative contributions from the charm and up quarks tend to cancel 
each other. To extract 
$g_{{D^{\ast}}^+ D^+ \gamma}$ reliablely, we combine the  
sum rules for $g_{{D^{\ast}}^+ D^+ \gamma}$ and $g_{{D^{\ast}}^0 D^0 \gamma}$ 
to get a new stable sum rules for the quantity ($g_{{D^{\ast}}^+ D^+ \gamma} +
g_{{D^{\ast}}^0 D^0 \gamma}$). Nevertheless, the error is larger for 
$g_{{D^{\ast}}^+ D^+ \gamma}$. 
The partial widths as calculated from the above couplings are given below. 
\begin{equation} 
\begin{array}{c}
\Gamma_{{D^{\ast}}^0\to D^0 \gamma} =12.9\pm 2 \mbox{keV}\, , \\
\Gamma_{{D^{\ast}}^+\to D^+ \gamma} =0.23\pm 0.1 \mbox{keV}\, , \\
\Gamma_{D_s^{\ast}\to D_s \gamma} =0.13\pm 0.05 \mbox{keV}\, ,
\end{array}
\end{equation}
which are also listed in Table I.

\par
Note that there are data from the CLEO collaboration \cite{cleo}:
\begin{equation}
\begin{array}{c}
 {\Gamma_{{D^{\ast}}^0 \to D^0 \gamma}  \over  \Gamma_{{D^{\ast}}^0 \to D^0 \pi^0}} 
=0.572\pm 0.057\pm 0.081 \, , \\
 {\Gamma_{\gamma}^+  \over  \Gamma_{\gamma}^0}=
 {\Gamma_{{D^{\ast}}^+ \to D^+ \gamma}  \over  \Gamma_{{D^{\ast}}^0 \to D^0 \gamma}}
=0.04 \pm 0.06\pm 0.06 \, ,     \\
B({D^{\ast}}^+ \to D^+ \gamma)=0.011\pm 0.014\pm 0.016 \, . 
\end{array}
\end{equation}
Our results yield 
\begin{equation}
\label{ratio}
 {\Gamma_{\gamma}^+  \over  \Gamma_{\gamma}^0}\approx 0.018  \, ,
\end{equation}
which is consistent with the CLEO data. On the other hand, the partial 
width for $D^{\ast}\to D +\pi$ has also been obtained making use of 
the light-cone QCD sum rules \cite{Be95}.
\begin{equation}
\label{p1}
\Gamma ({D^{\ast}}^0 \to D^0 \pi^0 ) =(23\pm 3) \mbox{keV} \, ,
\end{equation}
\begin{equation}
\label{p2}
\Gamma ({D^{\ast}}^+ \to D^0 \pi^+ ) +\Gamma ({D^{\ast}}^+ \to D^+ \pi^0 )
=(46\pm 7) \mbox{keV} \, .
\end{equation}
Combining our predictions with these results, we obtain
\begin{equation}
 {\Gamma_{{D^{\ast}}^0 \to D^0 \gamma}  \over  \Gamma_{{D^{\ast}}^0 \to D^0 \pi^0}} 
=0.56 \, , 
\end{equation}
\begin{equation}
B({D^{\ast}}^+ \to D^+ \gamma)= 0.005 \, ,
\end{equation}
both of which are in agreement with the CLEO data cited above.  

\newpage
\par
\noindent
TABLE I. Comparison between some theoretical calculations based on heavy 
quark and chiral symmetry \cite{Cho,Cheng}, the relativistic quark model \cite{JAUS}, 
and QCD sum rules approach (this work). The unit is keV.

\vskip 0.5 true cm
\begin{tabular}{|c|c|c|c|c|c|}
\hline
 & Cho$^a$ \cite{Cho} & Cho$^b$ \cite{Cho}& Cheng \cite{Cheng} &Jaus \cite{JAUS}& this work  \\
\hline
 ${D^{\ast}}^0\rightarrow D^0\gamma $ &$11.3\pm 22.0$&$8.8\pm 17.1$&$34$&$21.69$& $12.9\pm 2$ \\ 
 ${D^{\ast}}^+\rightarrow D^+\gamma $ &$10.6\pm 10.5$&$8.3\pm 8.1$&$2$&$0.56$& $0.23\pm 0.1$ \\ 
 ${D^{\ast}}_s\rightarrow D_s\gamma $ &-&-&$0.3$&-& $0.13\pm 0.05$ \\ 
 \hline
 ${B^{\ast}}^0\rightarrow B^0\gamma $ &$0.163\pm 0.261$&$0.127\pm 0.203$&$0.28$&$0.142$& $0.13\pm 0.03$ \\ 
 ${B^{\ast}}^+\rightarrow B^+\gamma $ &$0.846\pm 1.494$&$0.66\pm 0.931$&$0.84$&$0.429$& $0.38\pm 0.06$ \\ 
 ${B^{\ast}}_s\rightarrow B_s\gamma $ &-&-&-&-& $0.22\pm 0.04$ \\ 
\hline
\end{tabular}
\vskip 0.5 true cm
\noindent
$^a$ with the parameters $m_c=1500\,\mbox{MeV}$, $m_b=4500\,
\mbox{MeV}$. \\
$^b$ with the parameters $m_c=1700\,\mbox{MeV}$, $m_b=5000\,\mbox{MeV}$.
\vskip 0.5 true cm

\par
To sum up, we collect in Table I the existing theoretical predictions 
on the radiative decay widths for the bottomed and charmed mesons. As 
indicated earlier, our predictions in the charm sector are in good 
overall agreement with the existing CLEO data. In the bottom sector, 
there are differences (albeit less significant) among the various 
theoretical predictions but the data are not yet available.

\par
This work is supported in part by the National Natural Science Foundation
of China and by Doctoral Program Foundation of the State Commission. It is
also supported in part by the National Science Council of R.O.C. (Taiwan)
under the grant NSC86-2112-M002-010Y.

\end{document}